\documentclass[prl,preprintnumbers,amsmath,amssymb,nofootinbib,superscriptaddress,notitlepage,10pt]{revtex4-1}
\usepackage[utf8]{inputenc}
\usepackage[sort&compress]{natbib}
\usepackage{comment}
\usepackage{amssymb}
\usepackage{color}
\usepackage{rotating}
\usepackage{amsmath}
\usepackage{ulem}
\usepackage{overpic}
\usepackage{graphicx}
\usepackage{epstopdf}
\usepackage{dcolumn}
\usepackage{bm}
\usepackage{chngpage}
\usepackage{multirow}
\usepackage{slashed}
\usepackage{indentfirst}
\usepackage{booktabs}
\usepackage{amssymb,amsfonts}
\usepackage{amsmath}
\usepackage{color}
\usepackage{hyperref}

\begin{document}
\title{Excited $K$ meson, $K_c(4180)$, with hidden charm as a $D\bar{D}K$ bound state}
\author{Tian-Wei Wu}
\affiliation{School of Physics, Beihang University, Beijing 102206, China}
\author{Ming-Zhu Liu}
\affiliation{School of Space and Environment, Beihang University, Beijing 102206, China}
\author{Li-Sheng Geng\footnote{Corresponding author. lisheng.geng@buaa.edu.cn}}
\affiliation{School of Physics, Beihang University, Beijing 102206, China}
\affiliation{
Beijing Key Laboratory of Advanced Nuclear Materials and Physics,
Beihang University, Beijing 100191, China}
\affiliation{Beijing Advanced Innovation Center for Big Data-Based Precision Medicine, School of Medicine and Engineering, Beihang University, Beijing, 100191}
\affiliation{School of Physics and Microelectronics, Zhengzhou University, Zhengzhou, Henan 450001, China}

\date{September 2020}
\begin{abstract}
Motivated by the recent discovery of two new states in the $B^+\rightarrow D^+D^-K^+$ decay by the LHCb Collaboration,  we  study the $D\bar{D}K$ three-body system by solving the Schr\"odinger equation with the Gaussian Expansion Method. We show that the $D\bar{D}K$ system can bind with quantum numbers $I(J^P)=\frac{1}{2}(0^-)$ and a binding energy of $B_3(D\bar{D}K)=48.9^{+1.4}_{-2.4}$ MeV.  It can decay into $J/\psi K$ and $D_s\bar{D}^*$ via triangle diagrams, yielding a partial decay width of about 1 MeV. As a result, if discovered, it will serve as a highly nontrivial check on the nature of the many exotic hadrons discovered so far and on non-perturbative QCD as well. Assuming heavy quark spin symmetry, the same formalism is applied to study the $D\bar{D}^*K$ system, which is shown to also bind with quantum numbers $I(J^P)=\frac{1}{2}(1^-)$ and a binding energy of $B_3(D\bar{D}^*K)\simeq 77.3^{+3.1}_{-6.6}$ MeV, consistent with the results of previous works.

\end{abstract}
\maketitle

{\it Introduction:} Starting from 2003~\cite{Choi:2003ue,Aubert:2003fg}, many exotic hadronic states have been discovered experimentally. Most of them cannot easily fit into the conventional quark model picture, i.e., baryons consisting of $qqq$ and mesons of $q\bar{q}$. As a result, they have attracted a lot of attention and inspired intensive discussions about their true nature. Among the many possible interpretations, the  molecule picture has been  widely employed to interpret some of these exotic states (for  recent comprehensive reviews, see, e.g., Refs.~\cite{Hosaka:2016ypm,Esposito:2016noz,Chen:2016qju,Guo:2017jvc,Ali:2017jda,Olsen:2017bmm,Liu:2019zoy,Brambilla:2019esw}). Nonetheless,  it is  difficult if not impossible to distinguish different scenarios because  there are
always some free parameters in each model to fit
the experimental data. In this letter, we propose to study the existence of a three-body $D\bar{D}K$ bound state to decisively confirm or repute the molecular picture, with the assumption that $D_{s0}^*(2317)$ and
$D_{s1}(2460)$ are predominantly $DK$ and $D^*K$ bound states.

Being about 160 and 70 MeV lower than the corresponding $c\bar{s}$ states predicted by the naive quark model,   $D_{s0}^*(2317)$ and $D_{s1}(2460)$ can  be naturally interpreted as  $DK$  and $D^*K$ bound states~\cite{Kolomeitsev:2003ac,Guo:2006fu,Guo:2015dha,Geng:2010vw,Mohler:2013rwa}. If this is  the case, in the sense that the deuteron is a bound state of proton and neutron, a naive but straight forward question to ask is   whether $DDK$ (and/or $DD^*K$) can form  3-body bound states? A number of recent studies showed that they indeed bind~\cite{SanchezSanchez:2017xtl,MartinezTorres:2018zbl,Wu:2019vsy,Huang:2019qmw,Pang:2020pkl}. Lately, the Belle Collaboration has performed the first experimental search for the existence of the $DDK$ bound state and reported an upper limit for its production yield~\cite{Li:2020gvn}. 

Leading order chiral perturbation theory dictates that in the isospin zero channel, the $\bar{D}K$ and $\bar{D}^*K$ interactions are only  half those of the $DK$ and $D^*K$ interactions~\cite{Altenbuchinger:2013vwa}. In addition, both $X(3872)$ and $Z_c(3900)$ can be explained as $D\bar{D}^*$ molecules\footnote{More specifically, 
$X(3872)$ is a shallow bound state of $D\bar{D}^*$ with isospin zero~\cite{Sun:2011uh,Nieves:2012tt,Guo:2013sya,Karliner:2015ina} but $Z_c(3900)$ is a resonant state of $D\bar{D}^*$ with isospin 1~\cite{Aceti:2014uea,He:2015mja,Gong:2016hlt}.}, which implies that the $D\bar{D}^*$ interaction is attractive as well, but not as strong as the $DK$ and $D^*K$ interactions. A recent lattice QCD study  shows that the $D\bar{D}$ interaction is attractive such that a shallow $D\bar{D}$ bound state exists, consistent with the early theoretical results~\cite{Gamermann:2006nm} and more recent analysis of the $\gamma\gamma\rightarrow D\bar{D}$ reaction~\cite{Wang:2020elp}.
Motivated by these facts, the existence of a three-body $D\bar{D}^*K$ bound state has been studied.  In Ref.~\cite{Ma:2017ery}, the authors studied the $D\bar{D}^*K$ system using the Born-Oppenheimer approximation via delocalized $\pi$ bond. A bound state with the quantum numbers of $K^*$ and a mass of  $4317.92^{+6.13}_{-6.55}$ MeV was found. In Ref.~\cite{Ren:2018pcd}, using the so-called fixed-center approximation in coupled channels, the authors solved the Faddeev equation and found a heavy hidden charm $K^*$ meson  with a mass about $4307\pm2$ MeV, consistent with Ref.~\cite{Ma:2017ery}. 

Lately, the LHCb Collaboration found two new states in the $ B^+\rightarrow D^+D^-K^+$ decay~\cite{Aaij:2020hon,Aaij:2020ypa}, namely $X_0(2866)$ and $X_1(2900)$. The former has been interpreted as a $\bar{D}^*K^*$ bound state of spin zero~\cite{Liu:2020nil,Chen:2020aos,Hu:2020mxp,Huang:2020ptc,He:2020btl,Molina:2020hde,Agaev:2020nrc,Xiao:2020ltm}. Although a $D\bar{D}K$ bound state can not be found in the $D^+D^-K^+$ invariant mass spectrum, the recent experimental discovery indicates that a three-body $D\bar{D}K$ bound state, if it exists, could have been formed already and remain to be discovered at the current facilities. Motivated by these theoretical and experiment works, we study the strange hidden charm $D\bar{D}K$ system  using the Gaussian Expansion Method (GEM)~\cite{Hiyama:2003cu}. We indeed find a $D\bar{D}K$ bound state as well as  its heavy quark spin symmetry partner, a $D\bar{D}^*K$ bound state.

{\it Theoretical framework:}
\label{Sec:Framework}
There exit several widely used approaches to solve  three-body problems, such as the Faddeev equation~\cite{Yakubovsky:1966ue}, the Gaussian expansion method~\cite{Hiyama:2003cu}, the stochastic variational method~\cite{Varga:1995dm}, and the hyperspherical harmonic expansion method~\cite{Viviani:1994pm}. With the same inputs, the results of all these methods agree very well with each other as shown in  the benchmark study~\cite{Kamada:2001tv}. In this work, we utilize the Gaussian expansion method to study the $D\bar{D}^{(*)}K$ three-body systems, which has been widely used to solve three-, four- and even five-body  problems~\cite{Hiyama:2010zzd}, because of  its high precision and rapid convergence. Namely, we study the three-body $D\bar{D}^{(*)}K$ systems by solving the following Sch\"{o}dinger equation
\begin{equation}
\hat{H}\Psi=E\Psi,
\end{equation}
where the Hamiltonian $\hat{H}$ includes the kinetic term and three two-body interaction terms
\begin{equation}
  \hat{H}=T+V_{DK}+V_{\bar{D}^{(*)}K}+V_{D\bar{D}^{(*)}}.
\end{equation}
In order to solve the Sch\"{o}dinger equation, we have to first specify the two-body interactions.

For  the $DK$ interaction, we refer to  chiral perturbation theory, in which the most important contribution is the leading order Weinberg-Tomozawa (WT) term~\cite{Altenbuchinger:2013vwa}
\begin{equation}
    V_{DK}(\mathbf{q}) = -\frac{C_W(I)}{2 f_\pi^2},
\end{equation}
where the pion mass decay constant $f_{\pi}=130$ MeV and $C_{W}(I)$ represents the strength of the WT interaction with $C_{W}(0)=2$ for the isospin 0 and $C_{W}(1)=0$ for the isospin 1 configurations, respectively. This $DK$ potential can be rewritten in coordinate space by Fourier transformation and we use the same form of the $DK$ interaction as that adopted in Ref.~\cite{Wu:2019vsy}, which  explicitly reads
\begin{equation}
\label{Poten:DK}
    V_{DK}(r;R_c)=C(R_c)e^{-(r/R_c)^2}.
\end{equation}
Here $R_C$ is a coordinate space cutoff representing the effective interaction range. In this work, we choose $R_c$ ranging from 0.5 to 2.0 fm to study the related uncertainties. The $C(R_c)$ is a running constant related to $R_c$, which can be determined by reproducing the $D_{s0}^*(2317)$ state. 

The $\bar{D}K$ interaction can be related to the $DK$ interaction in chiral perturbation theory, where the leading order $\bar{D}K$ potential is half of the $DK$ interaction in the isospin zero channel~\cite{Altenbuchinger:2013vwa}. Thus we take the $\bar{D}K$ potential to have the same form as that of  Eq.~(\ref{Poten:DK}) but  multiplied with  1/2. Assuming heavy quark spin symmetry, the $\bar{D}^*K$ interaction is the same as the $\bar{D}K$ interaction.

%\begin{figure}[!h]
%  \centering
  % Requires \usepackage{graphicx}
% \includegraphics[scale=0.3]{DDbar.eps}
% \includegraphics[scale=0.3]{DDbarstar.eps}
%  \caption{The OBE potentials of $D\bar{D}$ (left) and $D\bar{D}^{*}$ (right) in S-wave.}
%  \label{DDbar}
%\end{figure}

For the $D\bar{D}$ interaction, there is no concrete experimental data and we have to resort to phenomenological models, e.g., the one boson exchange model of Ref.~\cite{Liu:2019stu}. In Ref.~\cite{Wu:2019vsy}, the $DD$ OBE potential has been derived with the exchange of $\sigma$, $\rho$ and $\omega$ mesons. According to G-parity, the only difference between the $D\bar{D}$ potential and the $DD$ potential in the OBE model is the sign of the $\omega$ exchange potential. The explicit form of the $DD$ interaction can be found in Ref.~\cite{Wu:2019vsy}.
For the $D\bar{D}^*$ interaction, one can also exchange a $\pi$ in addition to  the $\sigma$, $\rho$ and $\omega$ exchanges. 
We use the $D\bar{D}^*$ OBE potential of Ref.~\cite{Liu:2019stu} which reproduces the well-known $X(3872)$ state as a molecular state. We choose a cutoff $\Lambda=1.01$ GeV to reproduce the binding energy 4.0 MeV of $X(3872)$ with respect to the $D\bar{D}^{*}$ threshold. It should be mentioned that whether the $D\bar{D}$ system can form a bound state is still under discussion~\cite{Gamermann:2006nm,Prelovsek:2020eiw,Wang:2020elp}. In Ref.~\cite{Gamermann:2006nm}, the authors found a very narrow heavy scalar with mass around 3700 MeV (Re($\sqrt{s}$)=$3698\pm35$ MeV, Im($\sqrt{s}$)=$-0.10\pm0.06$ MeV). In Ref.~\cite{Prelovsek:2020eiw}, the authors found a shallow $D\bar{D}$ bound state with a binding energy $-4.0^{+3.7}_{-5.0}$ MeV in lattice QCD. In Ref.~\cite{Wang:2020elp}, the authors claimed that the $S$-wave $D\bar{D}$ final state interaction can produce a bound state around $3720$ MeV with $I=0$ by investigating the $\gamma\gamma \rightarrow D\bar{D}$ reaction.
In our model, with a cutoff $\Lambda=1.01$ GeV, the $D\bar{D}$ system can not form a bound state with the OBE potential, but will do so if a larger cutoff is adopted.~\footnote{More concretely, with a cutoff of 1.415 GeV, one can obtain a binding energy of 4 MeV for the $D\bar{D}$ system. The resulting $D\bar{D}K$ three-body bound state will have a binding energy of 20 MeV larger.} In the following, consistent with the $D\bar{D}^{*}$ case, we choose $\Lambda=1.01$ GeV for the $D\bar{D}$ OBE potential.

\begin{figure}[!h]
  \centering
  % Requires \usepackage{graphicx}
  \begin{overpic}[scale=0.5]{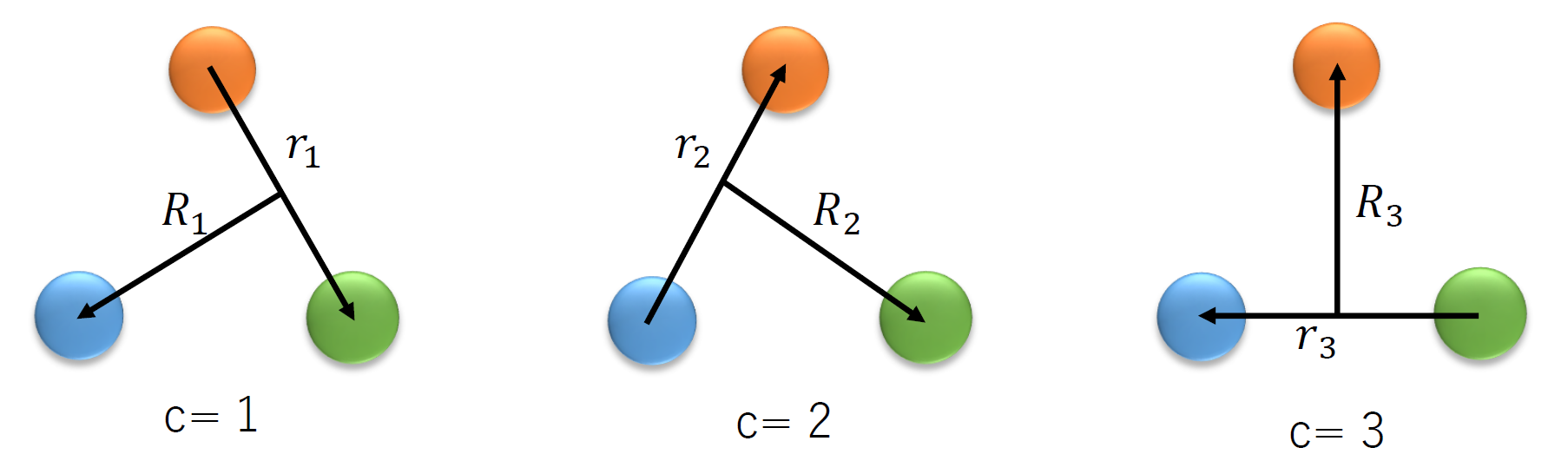}
  		\put(-3,9){$\bar{D}^{(*)}$}
  		\put(26,9){$D$}
  		\put(13,29){$K$}
		\put(33,9){$\bar{D}^{(*)}$}
		\put(62,9){$D$}
		\put(49,29){$K$}
		\put(68.5,9){$\bar{D}^{(*)}$}
		\put(98,9){$D$}
		\put(84.5,29){$K$}
  \end{overpic}
  \caption{Three permutations of the Jacobi coordinates for the $D\bar{D}^{(*)}K$ system.}
  \label{Jac}
\end{figure}
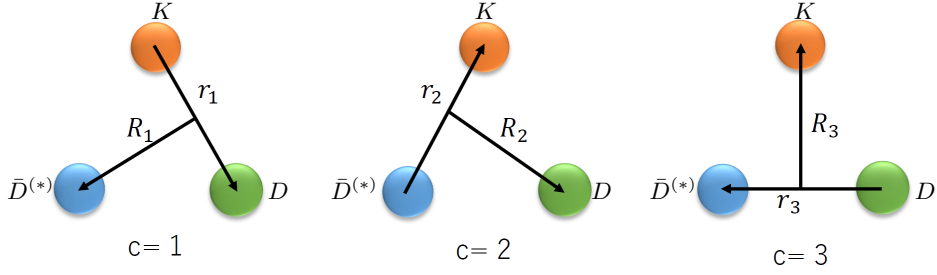

As all the two-body interactions have been specified, we use the GEM to solve the Sch\"odinger equation. The three-body wave functions can be constructed in Jacobi coordinates as 
\begin{equation}
    \Psi=\sum_{c=1}^{3}\Psi(\bf{r}_c,\bf{R}_c),
\end{equation}
where $c=1-3$ is the label of the Jacobi channels shown in Fig.~\ref{Jac}. In each Jacobi channel the wave function $\Psi(\mathbf{r}_c,\mathbf{R}_c)$ reads
\begin{equation}
    \Psi(r_c,R_c)=C_{c,\alpha}H^c_{t,T}\Phi_{lL,\lambda}(\mathbf{r}_c,\mathbf{R}_c)
\end{equation}
where $C_{c,\alpha}$ is the expansion coefficient and the $\alpha=\{nN,tT,lL\lambda\}$ labels the basis number with the configuration sets of the Jacobi channels. $H^c_{t,T}$ is the  three-body isospin wave function where $t$ is the isospin of the subsystem in Jacobi channel $c$ and $T$ is the total isospin. Considering that the isospin 1 $DK$ interaction is 0 in our model, the isospin $t_{DK}$ of the $DK$ subsystem should be 0 and thus the total isospin $T$ of the $D\bar{D}K$($D\bar{D}^*K$)  system is 1/2.

The three-body spatial wave function $\Phi(\mathbf{r}_c,\mathbf{R}_c)$ is constructed by two two-body wave functions as
\begin{equation}
\begin{split}
       \Phi_{lL,\Lambda}(\mathbf{r}_c,\mathbf{R}_c)&=[\phi_{n_cl_c}^{G}(\mathbf{r}_c)\psi_{N_cL_c}^{G}(\mathbf{R}_c)]_{\lambda},\\
     \phi_{nlm}^{G}(\mathbf{r}_c)&=N_{nl}r_c^le^{-\nu_n r_c^2} Y_{lm}({\hat{r}}_c),\\
     \psi_{NLM}^{G}(\mathbf{R}_c)&=N_{NL}R_c^Le^{-\lambda_n R_c^2} Y_{LM}({\hat{R}}_c).
\end{split}
\end{equation}
Here $N_{nl}(N_{NL})$ is the normalization constant of the Gaussian basis, $n(N)$  is the number of Gaussian basis used and $l(L)$ is the orbital angular momentum corresponding to the Jacobi coordinates $r (R)$.  Since only  $S$-wave interactions are considered, the total orbital angular momentum  is $\lambda=0$ and thus the quantum numbers $I
(J^P)$ of this  $D\bar{D}K$ three-body system are $\frac{1}{2}(0^-)$, while those for $D\bar{D}^*K$ are $\frac{1}{2}(1^-)$. 

With the constructed wave functions, the Sch\"odinger equation can be transformed into a generalized matrix eigenvalue problem with the Gaussian basis functions
\begin{equation}\label{eigenvalue problem}
  [T_{\alpha \alpha'}^{ab}+V_{{\alpha \alpha'}}^{ab}-EN_{\alpha \alpha'}^{ab}]\,
  C_{b,\alpha'} = 0
  \, ,
\end{equation}
where $T_{\alpha \alpha'}^{ab}$ is the matrix element of kinetic energy, $V_{\alpha \alpha'}^{ab}$ is the  matrix element of potential energy, and $N_{\alpha \alpha'}^{ab}$ is the normalization matrix element.

{\it Predictions and discussions:}
\label{Sec:Predictions}
We first study whether the three-body $D\bar{D}K$ and $D\bar{D}^*K$ systems bind in the  theoretical framework and with the two-body interactions specified  above.

The $D\bar{D}K$ system is found to bind with  quantum numbers $I(J^P)=\frac{1}{2}(0^-)$ and a three-body binding energy
\begin{equation*}
  B_3(D\bar{D}K)\simeq 48.9^{+1.4}_{-2.4}\quad\rm{MeV}.
\end{equation*}
 The results are weakly cutoff dependent, and therefore we vary the cutoff $R_c$ from $0.5$ to $2.0$ fm to estimate the uncertainties originating from the $DK$ and $\bar{D}K$ interactions. More concretely, the central value of the binding energy is obtained with $R_c=1.0$ fm while the uncertainties are taken from $R_c=0.5$ and 2.0 fm  in the specific numerical calculations.

Assuming heavy quark spin symmetry, the same formalism is applied to study the $D\bar{D}^*K$ system, where a bound state is found as well, with quantum numbers $I(J^P)=\frac{1}{2}(1^-)$ and a binding energy
\begin{equation*}
    B_3(D\bar{D}^*K)\simeq 77.3^{+3.1}_{-6.6}\quad\rm{MeV}.
\end{equation*}
The binding energy of the $D\bar{D}^*K$ bound state is larger than the one of $D\bar{D}^*K$ mainly due to the more attractive $D\bar{D}^*$ interaction.

\begin{table}[h]
    \caption{Masses and binding energies (in units of MeV) of the $D\bar{D}K$ and $D\bar{D}^*K$ bound states, in comparison with the results of other works.
    %Here, SE denotes Schr\"odinger equation; FE denotes Faddeev equation; GEM denotes Gaussian expansion method; BOA denotes Born-Oppenheimer approximation; FCA denotes fixed center approximation; $\chi$EFT denotes Chiral effective theory; and OBE denotes One boson exchange. 
    }
    \centering
    \begin{tabular}{c|c c c}
    \hline
    \hline
         & This work & Ref.~\cite{Ma:2017ery}& Ref.~\cite{Ren:2018pcd}\\
    \hline
    %   Method  &GEM(SE) & BOA(SE) & FCA(FE)\\
    %   Interaction Models& $\chi$EFT+OBE &delocalized $\pi$ bond &$\chi$EFT+OBE\\
       $\frac{1}{2}(0^-)$ $D\bar{D}K$&$4181.2^{+2.4}_{-1.4}$($B_3\simeq48.9^{+1.4}_{-2.4}$)  & - &-\\  
       $\frac{1}{2}(1^-)$ $D\bar{D}^*K$&$4294.1^{+6.6}_{-3.1}$($B_3\simeq77.3^{+3.1}_{-6.6}$)&$4317.92^{+6.13}_{-6.55}$($B_3\simeq53.52^{+6.55}_{-6.13}$) & $4307\pm2$($B_3\simeq64\pm2$)\\
    \hline\hline
    \end{tabular}
    \label{Results:BE}
\end{table}

As mentioned earlier, the $D\bar{D}^*K$ state has been studied using other methods. In Table~\ref{Results:BE}, we compare our predictions with those of two earlier studies.

It is clear that although there are some differences in detail, the existence of a $D\bar{D}^*K$ bound state seems to be a robust prediction.

\begin{figure}[!h]
  \centering
  % Requires \usepackage{graphicx}
 % \includegraphics[scale=0.5]{RMS1.png}\quad
 % \includegraphics[scale=0.5]{RMS2.png}\quad
 % \includegraphics[scale=0.5]{RMS3.png}
 \includegraphics[scale=0.5]{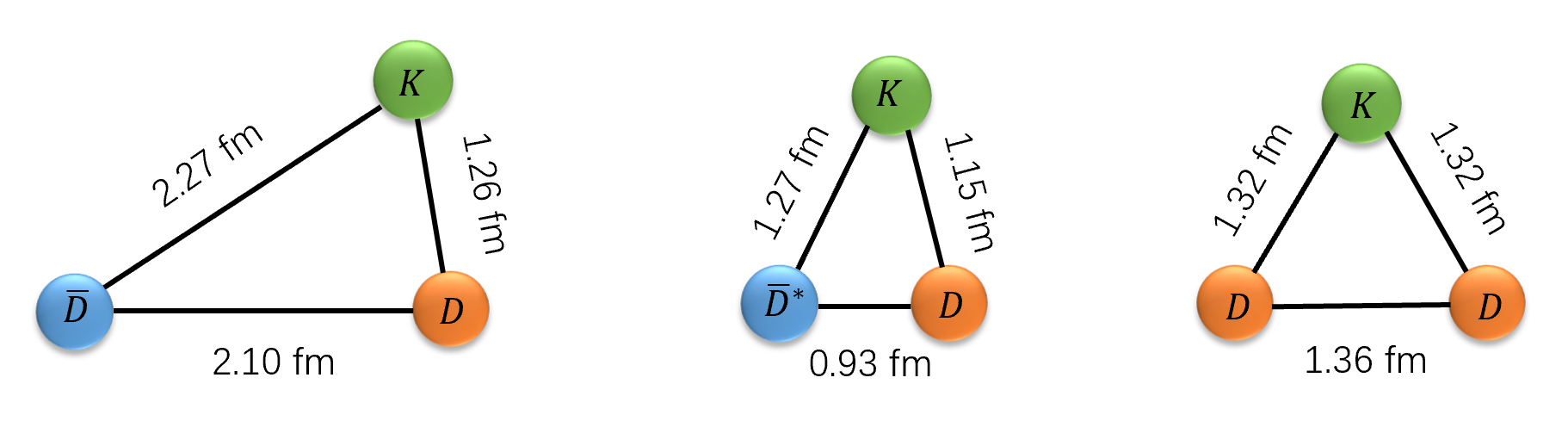}
  
  \caption{ RMS radii of subsystems in the $D\bar{D}K$ (left), $D\bar{D}^*K$ (middle), and $DDK$~\cite{Wu:2019vsy}  bound states with a cutoff $R_c=1.0$ fm.}
  \label{RMS}
\end{figure}

In Fig.~\ref{RMS}, we show the root mean square (RMS) radii of  the three subsystems in the $D\bar{D}K$ and $D\bar{D}^*K$ bound states. With a cutoff of $R_c=1.0$ fm, the RMS radius of the $DK$ subsystem in the $D\bar{D}K$ bound state is predicted to be $1.26$ fm,  while those of the $\bar{D}K$ and $D\bar{D}$ subsystems are much larger due to the less attractive interactions with respect to the $DK$ one, yielding $2.27$ and $2.10$ fm, respectively.
For the $D\bar{D}^*K$ bound state, the RMS radii for $DK$, $\bar{D}^*K$, and $D\bar{D}^*$  are $1.15$, 1.27, and 0.93 fm, respectively. The later two are much smaller compared to the ones of $\bar{D}K$ and $D\bar{D}$ in the $D\bar{D}K$ bound state. It can be easily understood because  the $D\bar{D}^*K$  bound state has a much larger binding energy. 
In Fig.~\ref{RMS}, we also show the RMS radii of the $DDK$ bound state taken from Ref.~\cite{Wu:2019vsy}, of which the $DK$ one is about $1.32$ fm and the $DD$ one is about $1.36$ fm.

\begin{figure}[!h]
\begin{center}
\begin{tabular}{cc}
\begin{minipage}[t]{0.4\linewidth}
\begin{center}
\begin{overpic}[scale=.6]{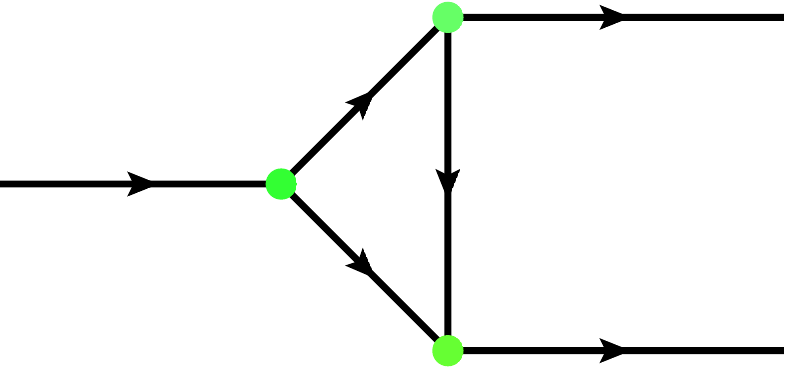}
		\put(74,6){$J/\psi$}
		
		\put(37,9){$\bar{D}$}
		
		\put(37,38){$D_{s0}^*$}
		
		\put(16,26){$K_{c}$ }
		\put(75,38){$K$} \put(60,22){$D$}
\end{overpic}
\end{center}
\end{minipage}
&
\begin{minipage}[t]{0.4\linewidth}
\begin{center}
%\subfigure[\tiny$\Phi_{s1}^{0}$]{
\begin{overpic}[scale=.6]{triangle.eps}
		\put(74,6){$\bar{D}^{\ast}$}
		
		\put(37,9){$\bar{D}$}
		
		\put(37,38){$D_{s0}^*$}
		
		\put(16,26){$K_{c}$ }
		\put(75,38){$D_{s}$} \put(60,22){$\eta$}
\end{overpic}
\end{center}
\end{minipage}
\end{tabular} \nonumber
\caption{Strong decays of $K_{c}(4180)$  to $J/\psi K$ and $D_{s}\bar{D}^{\ast}$ via triangle diagrams. \label{decay}}
\end{center}
\end{figure}
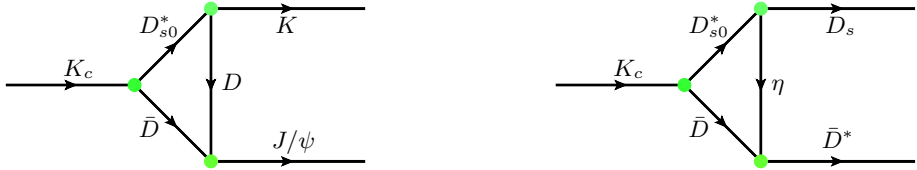

Following  Refs.~\cite{Huang:2019qmw,Ren:2019umd}  where the decays of the $DDK$ and $D^*\bar{D}K$ bound states via triangle diagrams have been estimated, in the following we compute the decay of the $D\bar{D}K$ state, which for convenience will be denoted by $K_c(4180)$. The decay of $K_c$ can proceed via its coupling to $D_{s0}^{\ast}$ and $\bar{D}$, and then decays to  $J/\psi K$ and $D_{s}\bar{D}^{\ast}$ via triangle diagrams  as shown in Fig.~\ref{decay}.  We  employ the effective Lagrangian approach to calculate the amplitudes of these hadronic loops.     

The interaction between $K_c(4180)$ and its components is described by the following effective Lagrangian
\begin{eqnarray}
\mathcal{L}_{K_c(x)}=g_{K_c D_{s0}\bar{D}}{K}_c^{T}(x)\int dy \Phi_{K_c}(y^2)D_{s0}(x+\omega_{\bar{D}}y)\bar{D}(x-\omega_{D_{s0}}y)+H.c.,
\end{eqnarray}
where $\omega_{i}=m_{i}/(m_{i}+m_{j})$ is a kinematical parameter with $m_{i}$ and $m_{j}$ being the masses of the components of $K_c$, $\Phi(y)$ is a form factor which we take a Gaussian form and  $g_{K_{c}D_{s0}\bar{D}}$ is the coupling constant between $K_c$ and its components. 
We can determine the value of $g_{K_c D_{s0}\bar{D}}$ by the compositeness condition, see, e.g.,  Refs.~\cite{Dong:2017gaw,Xiao:2019mvs,Huang:2019qmw}. In studies of hadronic molecules, the cutoff  in the form factor is often chosen to be about 1 GeV, and the corresponding coupling is found to be 5.38 GeV. The Lagrangians describing the interaction  between $D_{s0}^*(2317)$ and $DK$/$D_{s}\eta$ have the following form 
\begin{eqnarray}
\mathcal{L}_{D_{s0}DK}=g_{D_{s0}DK}D_{s0}DK,    \\ \nonumber
\mathcal{L}_{D_{s0}D_{s}\eta}=g_{D_{s0}D_{s}\eta}D_{s0}D_{s}\eta, 
\end{eqnarray}
where $g_{D_{s0}DK}$ and $g_{D_{s0}D_{s}\eta}$ denote the $D_{s0}^*(2317)$ coupling to $DK$ and $D_{s}\eta$   with $g_{D_{s0}DK}=10.21$ GeV and $g_{D_{s0}D_{s}\eta}=6.40$ GeV, respectively~\cite{Gamermann:2006nm}. The effective Lagrangian between  $J/\psi$ and $\bar{D}D$, and that between $\eta$ and $\bar{D}\bar{D}^{\ast}$  can be written as 
\begin{eqnarray}
\mathcal{L}_{\psi \bar{D}D}&=&-ig_{\psi \bar{D}D}\psi_{\mu}(\partial^{\mu}DD^{\dag}-D\partial^{\mu}D^{\dag}), 
\\ \nonumber
\mathcal{L}_{\bar{D}\bar{D}^{\ast}\eta}&=&-ig_{\bar{D}\bar{D}^{\ast}\eta}(\bar{D}^{\dag}\partial^{\mu}\eta \bar{D}^{\ast}-\bar{D}\partial_{\mu}\eta\bar{D}^{\ast\dag}),
\end{eqnarray}
where $g_{\psi \bar{D}D}={m_{\psi}}/f_{\psi}$ with $f_{\psi}=0.426$ GeV and $g_{\bar{D}\bar{D}^{\ast}\eta}=g/(\sqrt{3}f_{\pi})$ with $g=1.097$ GeV and $f_{\pi}=0.0924$ GeV \cite{Xiao:2019mvs}.   

With the above  Lagrangians, the amplitudes of the two triangle diagrams of Fig.~\ref{decay} can be straightforwardly calculated 
\begin{eqnarray}
i\mathcal{M}_{K_{c}\rightarrow J/\psi K}&=&g_{\psi \bar{D}D}g_{D_{s0}DK}g_{K_{c}D_{s0}\bar{D}}\int \frac{d^{4}q}{(2\pi)^{4}}\Phi[(k_{1}\omega_{\bar{D}}-k_{2}\omega_{D_{s0}})^{2}]\frac{1}{q^{2}-m_{D}^{2}}\frac{1}{k_{1}^{2}-m_{D_{s0}}^{2}}\frac{1}{k_{2}^{2}-m_{\bar{D}}^{2}}(q^{\mu}-k_{2}^{\mu})\cdot \varepsilon_{\mu}^{\psi},\nonumber \\ 
i\mathcal{M}_{K_{c}\rightarrow D_{s}\bar{D}^{\ast}}&=&g_{\bar{D}\bar{D}^{\ast}\eta}g_{D_{s0}D_{s}\eta}g_{K_{c}D_{s0}\bar{D}}\int \frac{d^{4}q}{(2\pi)^{4}}\Phi[(k_{1}\omega_{D}-k_{2}\omega_{D_{s0}})^{2}]\frac{1}{q^{2}-m_{\eta}^{2}}\frac{1}{k_{1}^{2}-m_{D_{s0}}^{2}}\frac{1}{k_{2}^{2}-m_{\bar{D}}^{2}}q^{\mu} \cdot \varepsilon_{\mu}^{\bar{D}^{\ast}},
\end{eqnarray}
where  $k_2$ and $k_1$ are the momenta of $\bar{D}$ and $D_{s0}^*(2317)$, $q$ the intermediate momenta, and $\epsilon_\mu^{\psi}$ and $\epsilon_\mu^{\bar{D}^*}$ are the polarization vectors of $\psi$ and $\bar{D}^*$.

With these amplitudes, the decay width can be easily obtained  
\begin{align}
\Gamma[K_{c}\to]=\frac{1}{2J+1}\frac{1}{32\pi^2}\frac{|\vec{p}_1|}{m_{K_{c}}^{2}}\overline{|{\cal{M}}|^2}d\Omega,
\end{align}
where $\vec{p}_1$ is the 3-momenta of either final state in the rest frame of $K_{c}$ and $m_{K_{c}}$ is the mass of the $D\bar{D}K$ bound state. 
With a cutoff  of 1 GeV for the form factor $\Phi_{K_c}(y^2)$, the decay width of $K_c(4180)$ to $J/\psi K$ and $D_{s}\bar{D}^{\ast}$ are found to be 0.5 MeV and 0.2 MeV, respectively, which indicates that this state is very narrow.
Compared to the $DDK$ bound state $R(4140)$, for which $\Gamma\simeq2-3$ MeV, the relative smallness of the decay width of $K_c(4180)$ is mainly due to the suppression of the $D$ exchange contribution in comparison with the corresponding $K$ exchange contribution.

{\it Summary:}
We employed the Gaussian expansion method to study the $D\bar{D}K$ system with the leading order $DK$ and $\bar{D}K$ potentials obtained in chiral perturbation theory and the $D\bar{D}$ potential from the OBE model. We found the existence of a $D\bar{D}K$ bound state with a binding energy about $49$ MeV. It is interesting to note that  the predicted $D\bar{D}K$ three-body bound state is only  4 MeV below the respective $\bar{D}D_{s0}^*(2317)$ threshold, though its binding could increase by 20 MeV if the $D\bar{D}$ interaction is strong enough to generate a bound state as claimed by the recent lattice QCD study.   We also studied its heavy quark spin partner, the $D\bar{D}^*K$ system, in the same framework, and we found a $D\bar{D}^*K$ bound state with a binding energy about of $77$ MeV,  consistent with the results of earlier works.

%To understand the spatial structure of the three-bound states, we calculated the root mean square radii of their two-body subsystems and compared them with those of the $DDK$ bound state. Our results are consistent with naive expectations, namely, 
%the more attractive the interactions, the deeper the bound states, and the   more compact their spatial distributions.

We studied the decays of the $D\bar{D}K$ bound state via triangle diagrams and found that  its partial decay widths to $J/\psi K$ and $D_s\bar{D}^*$ are about 0.5 MeV and 0.2 MeV, respectively.  Different from the $DDK$ bound state, these two exotic states are more likely to be discovered at the current facilities because of their hidden charm nature. It is interesting to note that a recent study in QCD sum rules does not find a $\bar{D}D_{s0}^*(2317)$ bound state~\cite{Di:2019qwv}, consistent with the current picture that the $K_c(4180)$ state is a three-body molecule. As a result, we strongly encourage our experimental colleagues to search for it.

{\it Acknowledgements:} This work was partly supported the National Natural Science Foundation of China (NSFC) under Grants Nos. 11975041, 11735003, and 11961141004.

\bibliography{DDbarK}

\end{document}